\documentclass{appolb}

\usepackage{epsfig}

\newcommand{\beq}{\begin{equation}}
\newcommand{\eeq}{\end{equation}}
\newcommand{\bi}{\begin{itemize}}
\newcommand{\ei}{\end{itemize}}
\newcommand{\beqar}{\begin{eqnarray}}
\newcommand{\eeqar}{\end{eqnarray}}

\let\epsilon\varepsilon

\newcommand{\pr}[1]{Phys. Rev. {\bf #1}}
\newcommand{\zp}[1]{Z. Phys. {\bf #1}}

\begin{document}

\title{THE MINIMAL EXTENSION OF THE SM AND THE NEUTRINO OSCILLATION DATA
\thanks{Presented by J. Gluza 
at the XXIII School of Theoretical Physics, Ustro\'n'99,
Poland, September 15-22, 1999.}$\;$\thanks{Work supported in part by 
Polish Committee for Scientific Research under 
Grants Nos. 2P03B08414 and 2P03B04215. 
J.G. would like to thank  also the Alexander von Humboldt-Stiftung for 
fellowship.}}
\author{ F. DEL AGUILA
\address {Departamento de Fisica Teorica y del Cosmos, \\
Universidad de Granada, 18071, Spain} \\ \vspace{.5 cm}
J. GLUZA
\address {Department of Field Theory and Particle Physics, Institute 
of Physics, \\ University of
Silesia, Uniwersytecka 4, PL-40-007 Katowice, Poland, \\
DESY Zeuthen, Platanenallee 6, 15738 Zeuthen, Germany} \\ \vspace{.5 cm}
M. ZRA\L EK
\address {Department of Field Theory and Particle Physics, Institute 
of Physics, \\ University of
Silesia, Uniwersytecka 4, PL-40-007 Katowice, Poland}
}

\maketitle

\begin{abstract}
We study the simplest Standard Model estension with only one extra 
right-handed neutrino. In this case there are two massless $m_{1,2}$ and 
two massive $m_{3,4}$ neutrinos, and in principle both solar and 
atmospheric 
anomalies can be described in two different scenarios, 
$m_3 \ll m_4$ (scheme I) and $m_3 \simeq m_4$ (scheme II). 
However, neither bi-maximal mixing nor the dark matter problem 
are explained in this minimal extension. 
Only scheme II can accommodate simultaneously maximal mixing for 
atmospheric neutrinos and the small mixing angle MSW solution for 
the solar anomaly. This scenario can be tested in the BOREXINO 
experiment.
\end{abstract}

\newpage

\section{Introduction}

The discovery of atmospheric muon neutrino oscillations by the large
Superkamiokande detector \cite{sup1} implies that neutrinos 
are massive particles. This  experiment has also strengthened 
the interpretation of the solar neutrino problem in terms of 
oscillation phenomena \cite{sup2}. 
The results of atmospheric neutrino experiments can be explained by $\nu
_\mu \rightarrow \nu _\tau $ oscillations with \cite{at}
\begin{equation}
\delta m_{Atm}^2\sim \left( 1.5-6 \right) \cdot 10^{-3}\; \ 
\mbox{\rm eV}^2\;\;\; \mbox{\rm and }\;\;\; A_{Atm}\sim 0.82-1.0 ,
\end{equation}
whereas solar neutrino experiments can be interpreted as result of the 
$\nu _e \rightarrow \nu _x\left(
x=\mu ,\tau \right) $ transition \cite{vo} with
\begin{equation}
\delta m_{sun}^2\sim \left( 0.5-0.8\right) \cdot 10^{-10} \ \ 
\mbox{\rm eV}^2, \mbox{ } A_{sun}\sim (0.72-0.95)
\end{equation}
in the case of vacuum oscillation (VO),
\begin{equation}
\delta m_{sun}^2\sim \left( 0.5-1\right) \cdot 10^{-5}\ \ eV^2,
\mbox{ } A_{sun}\sim (2-10) \cdot 10^{-3}
\end{equation}
in the case of small mixing angle (SMA) MSW transition \cite{msw1}, and  
\begin{equation}
\delta m_{sun}^2\sim \left( 0.16-4\right) \cdot 10^{-4} \ \ 
\mbox{\rm eV}^2,\mbox{ } A_{sun}\sim (0.65-1.0)
\end{equation}
in the case of large mixing angle (LMA) MSW transition.
Finally, let us also mention that the LSND data can be accommodated 
if \cite{lsnd}
\begin{equation}
\delta m_{LSND}^2\sim \left( 0.2-2\right) \ \mbox{\rm eV}^2\mbox{ and }
A_{LSND}\sim (0.3-4)\cdot 10^{-2} .
\end{equation}

There is  a vast literature exploring models of neutrino oscillations 
which can accommodate only two (atmospheric + solar) or all three 
(atmospheric + solar + LSND) anomalies. Most of them try
to understand why atmospheric and solar neutrino oscillations require
near maximal mixing (Eqs. (1) and (2)). Both are possible in the 
context of
three nearly--degenerate neutrinos or in see-saw models with
a neutrino mass hierarchy \cite{deg-see}. 
Scenarios with additional sterile neutrino(s) where all three anomalies 
can be explained are also investigated \cite{st}.

Here we study the simplest extension of the Standard Model (SM) 
with a single right-handed neutrino (RH1 model).   
Since the Higgs sector is not touched, the neutrino mass matrix
has four parameters. This simple matrix has two zero eigenvalues 
and we are not able to explain all three anomalies. Four different 
masses are needed to do that. 
So, we put the permanently unsettled LSND result aside and investigate 
in full detail solar and atmospheric anomalies in this model.
The diagonalization and mixing matrix obtained here can be used as a 
first 
step for diagonalizing the two (RH2) and three (RH3) right-handed 
neutrino SM extensions. There a full description of the neutrino data
will be possible \cite{fut}. 
  
\section{Model with one right-handed neutrino singlet}

In the SM with one Higgs doublet and one extra right-handed neutrino
singlet $\nu _{1R}$ the neutrino mass matrix has the form
\begin{equation}
M_\nu=\left( 
\begin{array}{cccc}
0 & 0 & 0 & a \\ 
0 & 0 & 0 & b \\ 
0 & 0 & 0 & c \\ 
a & b & c & M 
\end{array}
\right) 
\end{equation}
in the basis $\left( \nu _{eL},\nu _{\mu L},\nu _{\tau L},%
\overline{\nu }_{1L}^c\right) $.  
In this case CP is conserved \cite{ag} and all parameters can be assumed 
to be real and positive $\left( a,b,c,M\geq 0\right)$. The matrix 
$M_\nu$ is
diagonalized 
\begin{equation}
U^TM_\nu U = {\rm diag} \left( 0,0,m_3,m_4\right) 
\end{equation}
by the unitary transformation
\begin{equation}
U=\left[ 
\begin{array}{cccc}
s , & c \cos{\theta} , & ic\sin \theta \cos
\zeta , & c\sin \theta \sin \zeta  \\ 
-c  , & s \cos \theta , & is\sin \theta \cos
\zeta , & s\sin \theta \sin \zeta  \\ 
0 , & -\sin \theta , & i\cos \theta \cos \zeta , & \cos
\theta \sin \zeta  \\ 
0, & 0, & -i\sin \zeta , & \cos \zeta 
\end{array}
\right] ,
\end{equation}
where
\begin{eqnarray}
m_{3,4}&=&\frac 12M\left[ \sqrt{1+4 \left( \frac \Lambda M \right)^2}
\mp 1\right] , 
 \\
\sin \theta &=&\frac \lambda \Lambda ,\;\;\;\; 
\cos \theta =\frac c\Lambda , 
 \\
s\equiv \sin \varphi &=&\frac b\lambda ,\;\;\;\; c\equiv 
\cos \varphi =\frac a\lambda ,
\\
\sin \zeta &=&\sqrt{\frac{m_3}{m_3+m_4}},\;\;\;\; 
\cos \zeta =\sqrt{\frac{m_4}{%
m_3+m_4}}, 
\\
\lambda &=& \sqrt{a^2+b^2},\;\;\;\; \Lambda =\sqrt{a^2+b^2+c^2}. 
\end{eqnarray}
The two massive neutrinos have opposite CP parities and  
the non-zero masses are function only of  
M and $\Lambda/M$. If $\Lambda \ll M$, the traditional see-saw mechanism
works. This case  with $M$ greater than 1 GeV or even than $M_Z$ 
(heavy neutrino singlet) has been discussed in \cite{jar,ng,esc}. 
$m_4$ is then $\sim M$.  
However, we are not interested in such a case since we need much smaller 
$m_{3,4}$ masses to be able to explain simultaneously the small mass 
squared splittings dictated by solar and atmospheric results. 
Two different scenarios are possible in this simple model, 
in scheme I  $m_3 \ll m_4$ and in scheme II $m_3 \simeq m_4$ (Fig.(1)).
\begin{figure}
\epsfig{figure=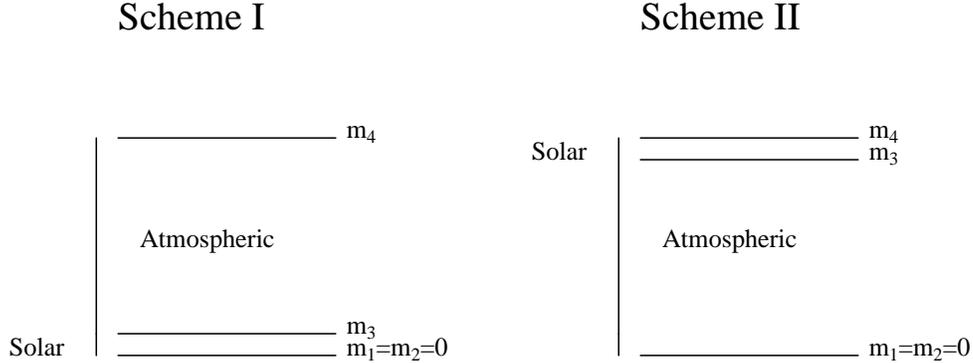, height=2 in}
\caption{Two possible neutrino mass spectra which can describe the 
oscillation data in the RH1 model.}
\end{figure}
Since two masses are zero, the absolute scale of the neutrino mass 
spectrum 
is constrained and $m_{3,4}$ fixed, 
in contrast with the general case where all neutrinos can be massive
\cite{bar,cza}. Eq. (1) requires 
\begin{equation}
0.038 \; {\rm eV}  \leq m_4 \leq 0.078 \; {\rm eV}
\end{equation}
in both schemes. Once $m_{1,2,4}$ are determined, $m_3$ is fixed by 
$\delta m_{sun}^2$ (Eqs. (2-4)).
Hence, we are really interested in quite small $M$ ($\Lambda$) values  
ranging in the milielectronvolt scale. 
Let us note that these masses do not solve the dark matter problem.
Two further remarks are in order. 
First, the smallness of the neutrino masses compared to other known
particles implies no problem with non-oscillation experiments. 
For example, the number of neutrino species measured by 
LEP1 is  $N_\nu=3$ (four neutrinos are produced in $Z$ decay)
\cite{jar,ng,esc}.
Second, the Heidelberg-Moscow limit on the effective neutrino mass, 
$<m_{\nu}>_{ee} \ \equiv \ U_{ei}^2\ m_i \leq 0.2 $ eV 
from the non-observation of the neutrinoless double beta decay 
\cite{beta}, 
is automatically fulfilled. 
$<m_{\nu}>_{ee}$ is the element (1,1) of $M_{\nu}$ in Eq. (6) 
which is equal to zero. 

\section{Oscillation probabilities and study of the model}

Let us apply the probability of the flavour changing 
$\alpha \rightarrow \beta $ neutrino transition in vacuum, which is a 
function of the travelling distance L,
\begin{equation}
P_{\alpha \rightarrow \beta }(L)=\delta _{\alpha \beta }-
\sum_{a>b}\left( 4R_{\alpha \beta }^{ab}
\sin ^2\Delta _{ab}-2I_{\alpha \beta
}^{ab}\sin 2\Delta _{ab}\right) 
\end{equation}
where
$$
R_{\alpha \beta }^{ab}=Re\left[ U_{\alpha a}U_{\beta b}U_{\alpha
b}^{*}U_{\beta a }^{*}\right] , \;\;\;\;\;\;\;\;
I_{\alpha \beta }^{ab}=Im\left[ U_{\alpha a}U_{\beta b}U_{\alpha
b}^{*}U_{\beta a }^{*}\right] , 
$$
and
$$
\Delta _{ab}=1.27\delta m_{ab}^2\left( 
{\rm eV}^2\right) \frac{L\left( {\rm km}\right) }
{E\left( {\rm GeV}\right) }, 
$$
to both mass spectra. 

\subsection{Scheme I}

In this case the oscillation probability reads 
\begin{equation}
P_{\alpha \rightarrow \beta } \simeq \delta _{\alpha \beta }-\left(
4(R_{\alpha \beta }^{41}+R_{\alpha \beta }^{42}+
R_{\alpha \beta }^{43})\sin
^2\Delta _{atm}+4(R_{\alpha \beta }^{31}+R_{\alpha \beta }^{32})\sin
^2\Delta _{sun}\right), 
\label{osc}
\end{equation}
where
$$
\Delta _{atm}\simeq \Delta _{43}\simeq \Delta _{41}=\Delta _{42} \;\;\;
\mbox{and} \;\;\;\;
\Delta _{sun}\simeq \Delta _{31}=\Delta _{32}. 
$$
For $L= L_{atm}= \left( 20 \div 10^4\right) $ km, $\Delta
_{atm}\gg \Delta _{sun}$ and  the second  oscillation term in 
Eq. (\ref{osc}) has no time to develop.
The oscillation of atmospheric neutrinos is then described by 
\begin{equation}
P_{\alpha \rightarrow \beta }\left( L_{atm}\right) \simeq 
\delta _{\alpha
\beta }-4(R_{\alpha \beta }^{41}+R_{\alpha \beta }^{42}+
R_{\alpha \beta
}^{43})\sin ^2\Delta _{atm}.
\end{equation}
On the other hand, 
at the solar distance scale $L=L_{solar}\sim 10^8$ km the first 
oscillation term
is averaged, $ \sin ^2\Delta _{atm}\rightarrow \frac 12 $, 
and the flavour changing probability is
\begin{equation}
P_{\alpha \rightarrow \beta }\left( L_{solar}\right) \simeq 
\delta _{\alpha
\beta }-2(R_{\alpha \beta }^{41}+R_{\alpha \beta }^{42}+
R_{\alpha \beta
}^{43})-4(R_{\alpha \beta }^{31}+R_{\alpha \beta }^{32})
\sin ^2\Delta _{sun}.
\end{equation}
Now, it is straightforward to find the relevant oscillation 
probabilities. 
For atmospheric neutrinos we have: 
\begin{eqnarray}
P_{\nu _\mu \rightarrow \nu _e}\left( L_{atm}\right) &\simeq&
\sin ^22\varphi \sin
^4\zeta \sin ^4\theta \sin ^2\Delta _{atm}, \label{mue}
\\
P_{\nu _\mu \rightarrow \nu _\tau }\left( L_{atm}\right)&\simeq&
\sin ^2\varphi
\sin ^4\zeta \sin ^22\theta \sin ^2\Delta _{atm},
\\
P_{\nu _\mu \rightarrow \nu _s}\left( L_{atm}\right) &\simeq&
\sin ^2\varphi \sin
^22\zeta \sin^2 \theta \sin ^2\Delta _{atm},
\end{eqnarray}
and 
\begin{eqnarray}
P_{\nu _\mu \rightarrow \nu _\mu}\left( L_{atm}\right)& \simeq&
1-4\sin^2{\varphi} \sin
^2\zeta \sin^2\theta  \label{mumu}  \\
& \times & \left( \cos ^2\varphi \sin ^2\zeta \sin ^2\theta
+\sin ^2\zeta \cos ^2\theta +\cos ^2\zeta \right) \sin ^2\Delta
_{atm}. \nonumber
\end{eqnarray}
Whereas for solar neutrinos the oscillation probabilities are: 
\begin{eqnarray}
P_{\nu _e\rightarrow \nu _\mu }\left( L_{solar}\right)& \simeq&
\sin ^22\varphi \sin
^4\theta \left( \frac 12\sin ^4\zeta +\cos ^2\zeta \sin ^2\Delta
_{sun}\right) , \label{emu} \\
P_{\nu _e\rightarrow \nu _\tau }\left( L_{solar}\right) & \simeq & \cos
^2\varphi \sin ^22\theta \left( \frac 12\sin ^4\zeta +\cos ^2\zeta
\sin ^2\Delta _{sun}\right) , \label{etau}\\
P_{\nu _e\rightarrow \nu _s}\left( L_{solar}\right) & \simeq & \frac 
12\cos
^2\varphi \sin ^22\zeta \sin ^2\theta , 
\end{eqnarray}
and
\begin{eqnarray}
P_{\nu _e\rightarrow \nu _e}\left( L_{solar}\right) & \simeq & 
1-2\cos ^2\varphi
\sin ^2\theta  \nonumber \\ 
&&\left[ \sin ^2\zeta \left( \sin ^2\varphi \sin ^2\zeta \sin
^2\theta +\sin ^2\zeta \cos ^2\theta +\cos ^2\zeta \right) \right. 
\nonumber  \\
&+& \left. 2\cos
^2\zeta \left( \sin ^2\varphi \sin ^2\theta +\cos ^2\theta \right)
 \sin ^2\Delta _{sun} \right].
\end{eqnarray}
Since  $m_4\gg m_3$, 
\begin{equation}
\sin \zeta \ll \cos \zeta \sim 1.
\end{equation}
The oscillation parts of $P_{\nu _e\rightarrow \nu _\mu }$ 
and $P_{\nu _e\rightarrow \nu _\tau }$ for solar neutrinos, 
Eqs. (\ref{emu}) and (\ref{etau}) respectively, are
proportional to cos$^2\zeta $ . Depending on the angles $\varphi $ and
$\theta$, the mixing can be large or small, so any solution 
(Eq. (2), (3)
or (4)) is possible. Unfortunately, the 
probabilities for atmospheric neutrinos (Eqs. (\ref{mue}--\ref{mumu})) 
are proportional to $\sin ^2\zeta $ and very small. Then, it is 
impossible 
to explain the observed atmospheric
neutrino anomaly in this scheme.

\subsection{Scheme II}

In this case $m_3 \simeq m_4$ and 
\begin{equation}
P_{\alpha \rightarrow \beta }\left( L_{atm}\right)  \simeq 
\delta _{\alpha
\beta }-4\left( R_{\alpha \beta }^{31}+R_{\alpha \beta }^{32}+R_{\alpha
\beta }^{41}+R_{\alpha \beta }^{42}\right) \sin ^2\Delta _{atm}
\end{equation}
for atmospheric neutrino oscillations, and 
\begin{equation}
P_{\alpha \rightarrow \beta }\left( L_{solar}\right) \simeq 
\delta _{\alpha
\beta }-2\left( R_{\alpha \beta }^{31}+R_{\alpha \beta }^{32}+R_{\alpha
\beta }^{41}+R_{\alpha \beta }^{42}\right) -4R_{\alpha \beta }^{43}\sin
^2\Delta _{sun}
\end{equation}
for solar neutrino oscillations. These probabilities reduce for the 
specific 
transitions to
\begin{eqnarray}
P_{\nu _\mu \rightarrow \nu _e}\left( L_{atm}\right)& \simeq& 
\sin ^22\varphi \sin
^4\theta \sin ^2\Delta _{atm},\\
P_{\nu _\mu \rightarrow \nu _\tau }\left( L_{atm}\right) & \simeq& 
\sin ^2\varphi
\sin ^22\theta \sin ^2\Delta _{atm},\\
P_{\nu _\mu \rightarrow \nu _s}\left( L_{atm}\right) & \simeq&0,\\ 
P_{\nu _e\rightarrow \nu _\mu }\left( L_{solar}\right)&  \simeq&
\frac 12\sin
^22\varphi \sin ^4\theta \left( 1-\frac 12\sin ^22\zeta 
\sin ^2\Delta
_{sun}\right) ,\\
P_{\nu _e\rightarrow \nu _\tau }\left( L_{solar}\right) & \simeq&
\frac 12\cos
^2\varphi \sin ^22\theta \left( 1-\frac 12\sin ^22\zeta \sin ^2\Delta
_{sun}\right) ,\\
P_{\nu _e\rightarrow \nu _s}\left( L_{solar}\right) & \simeq& 
\cos ^2\varphi \sin
^22\zeta  \sin ^2\theta \sin ^2\Delta _{sun} 
\end{eqnarray}
and
\begin{eqnarray}
P_{\nu _\mu \rightarrow \nu _\mu }\left( L_{atm}\right) &  \simeq& 
1-\left( \sin
^22\varphi \sin ^4\theta +\sin ^2\varphi \sin ^22\theta  \right) 
\sin ^2\Delta
_{atm} , \nonumber \\
&& \label{mumusup}\\
P_{\nu _e\rightarrow \nu _e}\left( L_{atm(CHOOZ)}\right)&  \simeq&
1-\left( \cos ^2\varphi
\sin ^22\theta +\sin ^22\varphi \sin ^4\theta \right) 
\sin ^2\Delta_{atm} ,
\nonumber \\
&& \label{eech}\\
P_{\nu _e\rightarrow \nu _e}\left( L_{solar}\right) & \simeq& 
1-\frac 12(\sin ^22\varphi \sin ^4\theta +\cos ^2\varphi 
\sin^22\theta ) 
\nonumber \\ &-&\cos
^4\varphi \sin ^22\zeta \sin ^4\theta \sin ^2\Delta _{sun}\label{eesol}.
\end{eqnarray}
Since the non-zero masses are nearly degenerate, 
the mixing angle $\zeta$ is almost maximal
\begin{equation}
\sin \zeta \simeq \cos \zeta \sim \frac 1{\sqrt{2}}.
\end{equation}
The CHOOZ reactor experiment \cite{chooz} constrains 
$P_{\nu _e\rightarrow \nu _e}$ (Eq. (\ref{eech})), 
\begin{equation}
\cos ^2\varphi \sin ^22\theta +\sin ^22\varphi \sin ^4\theta 
\leq 0.18 \;\; \mbox{for }\delta m^2>0.9 \cdot 10^{-3} \; eV^2,
\label{chooz}
\end{equation}
and the Superkamiokande experiment 
$P_{\nu _\mu \rightarrow \nu _\mu }$ (Eq. (\ref{mumusup})),  
\begin{equation}
0.82\leq \sin ^22\varphi \sin ^4\theta +\sin ^2\varphi \sin ^22\theta
\leq 1.
\label{sup}
\end{equation}
Both restrictions are satisfied if  
cos$\varphi \sim 0$ and sin$2\theta \sim 1$. 
However, in this case the solar neutrinos do not oscillate (Eq.
(\ref{eesol})). This means that bi-maximal mixing for solar and 
atmospheric neutrinos is not possible in the RH1 model.
Although recent Superkamiokande data favour vacuum long-wavelength 
oscillation 
of solar neutrinos, this can not be
explained with only one extra right-handed neutrino singlet. 
However, the deficit of solar neutrinos can be also described by 
the SMA MSW 
transition (Eq. (3)) and all present observations 
(without LSND data) can be then accommodated in this minimal SM 
extension.
Indeed the CHOOZ (Eq. (\ref{chooz})) and 
Superkamiokande (Eq. (\ref{sup})) constraints are also fulfilled 
if $\cos \varphi \gg 0$ and sin$2\theta \ll 1$. In this case   
(see Eq. (\ref{eesol}))
\begin{equation}
A_{sun} \simeq \cos^4 \varphi \sin ^4\theta 
\end{equation}
satisfies Eq. (3). 
For example, $\cos^2 \varphi = 0.17$ and
sin$^2\theta = 0.35$ verify Eqs. (\ref{chooz}) and (\ref{sup}), 
implying 
\begin{equation}
A_{sun}=0.0035
\end{equation}
which lies within the SMA MSW limits. 
The mixing angles determine the mixing matrix in Eq.(8) 
\begin{eqnarray}
\nu _e&=&+0.91\nu _1+0.33\nu _2+i0.17\nu _3+0.17\nu _4, 
\nonumber \\
\nu _\mu & =& -0.41\nu _1+0.73\nu _2+i0.38\nu _3+0.38\nu _4,
\\
\nu _\tau &=&-0.59\nu _2+i0.57\nu _3+0.57\nu _4, 
\nonumber \\ 
\nu _s&=&-i0.71\nu _3+0.71\nu _4, \nonumber 
\end{eqnarray}
and Eqs. (1) and (3) are fulfilled by the neutrino masses  
\begin{equation}
m_3=0.05477\; \mbox{\rm eV},\; \ \;m_4=0.05482\; \mbox{\rm eV}.
\end{equation}
These eigenvectors and eigenvalues are obtained from the $M_\nu $
entries (Eq. (6))
\begin{eqnarray}
a = 0.013376\ \mbox{\rm eV}, b = 0.02953\ \mbox{\rm eV}, c = 0.04418 
\ \mbox{\rm eV}, M = 5 \cdot 10^{-5} \ \mbox{\rm eV}.
\end{eqnarray} 
In this model contrary to what happens in the popular see-saw mechanism, 
the right-handed
Majorana mass term $M$ is much smaller than the Dirac masses $a, b, c$.

\section{Conclusions}

The RH1 model seems to be too simple to explain the observed 
neutrino anomalies. 
The popular bi-maximal solution for the atmospheric and solar 
anomalies can not
be realized in this model, neither the dark matter problem can 
be solved.
Although not favoured, only the small mixing angle MSW transition 
for solar neutrinos and the maximal neutrino mixing oscillation 
solution 
for atmospheric neutrinos can be accommodated. 
The model which is the simplest SM extension, will be definitively 
excluded
if the favoured `just so' mechanism for solar neutrinos persists.

\end{document}